\documentclass[conference]{IEEEtran}
\IEEEoverridecommandlockouts
\usepackage{cite}
\usepackage{amsmath,amssymb,amsfonts}
\usepackage{algorithmic}
\usepackage{graphicx}
\usepackage{textcomp}
\usepackage{xcolor}

\usepackage{graphicx}

\usepackage{amsmath}
\usepackage{amsfonts, amssymb}
\usepackage{booktabs}
\usepackage{tabularx}
\usepackage{multirow}
\usepackage[shortcuts]{extdash}

\def\BibTeX{{\rm B\kern-.05em{\sc i\kern-.025em b}\kern-.08em
    T\kern-.1667em\lower.7ex\hbox{E}\kern-.125emX}}
\begin{document}

\title{MITIA: Reliable Multi-modal Medical Image-to-image Translation Independent of Pixel-wise Aligned Data\\
    \thanks{
        This paper has been accepted as a research article by Medical Physics. Manuscript received 11 May 2024; revised 19 July 2024; accepted 4 August 2024. This work is supported in parts by the National Key Research and Development Program of China (2022YFF0710800), and Jiangsu Provincial Key Research and Development Program (BE2021609).
        Langrui Zhou and Guang Li are with Jiangsu Key Laboratory for Biomaterials and Devices, School of Biological Science and Medical Engineering, Southeast University, Nanjing, China.\\
        \indent Correspondence: liguang@seu.edu.cn.
        Langrui Zhou and Guang Li contributed equally to this study.\\
        \indent Digital Objective Identifier: 10.1002/mp.17362
    }
}

\author{\IEEEauthorblockN{Langrui Zhou}
        \IEEEauthorblockA{Southeast University}
        \and
        \IEEEauthorblockN{Guang Li}
        \IEEEauthorblockA{Southeast University}
}

\maketitle

\begin{abstract}
The current mainstream multi-modal medical image-to-image translation methods face a contradiction. Supervised methods with outstanding performance rely on pixel-wise aligned training data to constrain the model optimization. However, obtaining pixel-wise aligned multi-modal medical image datasets is challenging. Unsupervised methods can be trained without paired data, but their reliability cannot be guaranteed. At present, there is no ideal multi-modal medical image-to-image translation method that can generate reliable translation results without the need for pixel-wise aligned data.
This work aims to develop a novel medical image-to-image translation model that is independent of pixel-wise aligned data (MITIA), enabling reliable multi-modal medical image-to-image translation under the condition of misaligned training data.
The proposed MITIA model utilizes a prior extraction network composed of a multi-modal medical image registration module and a multi-modal misalignment error detection module to extract pixel-level prior information from training data with misalignment errors to the largest extent. The extracted prior information is then used to construct a regularization term to constrain the optimization of the unsupervised cycle-consistent GAN model, restricting its solution space and thereby improving the performance and reliability of the generator. We trained the MITIA model using six datasets containing different misalignment errors and two well-aligned datasets. Subsequently, we conducted quantitative analysis using peak signal-to-noise ratio (PSNR) and structural similarity (SSIM) as metrics. Moreover, we compared the proposed method with six other state-of-the-art image-to-image translation methods.
The results of both quantitative analysis and qualitative visual inspection indicate that MITIA achieves superior performance compared to the competing state-of-the-art methods, both on misaligned data and aligned data. Furthermore, MITIA shows more stability in the presence of misalignment errors in the training data, regardless of their severity or type.
The proposed method achieves outstanding performance in multi-modal medical image-to-image translation tasks without aligned training data. Due to the difficulty in obtaining pixel-wise aligned data for medical image translation tasks, MITIA is expected to generate significant application value in this scenario compared to existing methods.
\end{abstract}


\section{Introduction}
Multi-modal medical imaging is crucial for improving diagnostic accuracy\cite{b1}. 
However, acquiring multi-modal medical images often involves high economic and labor costs\cite{b2,b3,b4}.
To facilitate the acquisition of multi-modal medical images, deep learning-based multi-modal medical image-to-image translation methods have been widely proposed\cite{b5,b6,b7,b8}. 
Among them, the most representative method is the Generative Adversarial Network (GAN)\cite{b9}.
After years of continuous development, GAN has become one of the most commonly used methods in the field of image-to-image translation\cite{b10,b11,b12}.
GAN-based image-to-image translation methods can be divided into supervised and unsupervised methods.
Supervised methods\cite{b6,b13} optimize the generator by minimizing pixel-wise loss between the predicted image $G(x)$ and the target image $y$.
Since the training data is pixel-wise aligned, each pixel in the source domain image has a corresponding label in the target domain image.
Therefore, generators trained based on supervised methods can predict reliable and high-quality translation results.
However, in medical scenarios, collecting pixel-wise aligned datasets is very expensive, time-consuming, and often impossible to achieve in many cases, which greatly limits the applicability of supervised methods in multi-modal medical image-to-image translation tasks.
To overcome the limitations of pixel-wise aligned data, unsupervised methods, primarily based on cycle-consistency constraints, have been widely proposed\cite{b14,b15,b16}.
By adding a reverse generator $F:Y\rightarrow X$ to complete the inverse mapping of $G:X\rightarrow Y$ and introducing cycle-consistency loss to enforce $F(G(X))\approx X$ and $G(F(Y))\approx Y$, unsupervised methods avoid pixel-wise cross-domain loss and can achieve excellent performance without paired data.
However, unsupervised methods still have their shortcomings in medical image-to-image translation tasks.
In medical images, each anatomical structure has a strictly defined range of pixel values.
To ensure that the translated results retain as much anatomical information as possible, medical image-to-image translation tasks should have a unique optimal solution.
However, in practical applications, there are often multiple mappings between the source domain and the target domain that satisfy the cycle-consistency constraints.
Therefore, unsupervised methods may suffer from the ``multiple solutions'' problem\cite{b17,b18}, which is unacceptable in medical image-to-image translation tasks.
Recently, although researchers have attempted to use novel methods other than GANs, such as diffusion models\cite{b7,b19,b20}, to achieve performance improvements in image-to-image translation tasks, the contradiction between the reliability of translation results and the accessibility of training data has not been effectively resolved.
To this day, there is still no ideal medical image-to-image translation method that can achieve outstanding performance without the need for pixel-wise aligned training data.

In multi-modal medical image-to-image translation tasks, using pixel-level prior information in training data to constrain the model optimization is crucial for enhancing the performance and reliability of the generator. 
Unsupervised methods suffer from an ambiguous solution space due to a lack of pixel-wise prior constraints. Supervised methods perform well, but the introduction of misalignment errors in the preparation of multi-modal medical image data is almost unavoidable.
Using data with misalignment errors for training would negatively impact the performance of supervised methods. In fact, for multi-modal medical images of the same sample, although there are often misalignment errors between these images, they still contain abundant available pixel-level prior information. 
If these pixel-level prior information can be appropriately extracted from the misaligned data and used to constrain the model optimization, it will be possible to make reliable multi-modal medical image-to-image translation independent of pixel-wise aligned data.
Based on this idea, in this paper, we propose MITIA, a multi-modal medical image-to-image translation model that does not rely on pixel-wise aligned data.
MITIA utilizes a prior extraction network composed of a multi-modal medical image registration module and a multi-modal misalignment error detection module to extract pixel-level prior information from training data with misalignment errors to the largest extent. The extracted prior information is then used to construct a regularization term to constrain the optimization of the unsupervised cycle-consistent GAN model, restricting its solution space and thereby improving the performance and reliability of the generator.

The remainder of this paper is organized as follows. In Section II, we first briefly analyze the misalignment errors in multi-modal medical images, and then elaborate on the proposed MITIA model. In Section III, we validate the performance of MITIA using six misaligned datasets and two well-aligned datasets, respectively. In Section IV, results and relevant issues are discussed, and the conclusions are drawn.

\section{Methodology}

\subsection{Motivation}
\begin{figure}[!htb]
    \centerline{\includegraphics{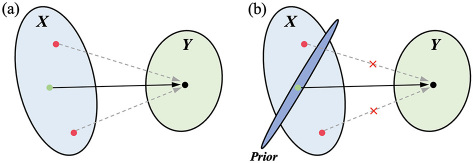}}
    \caption{(a) Unsupervised cycle-consistent methods may produce multiple solutions. (b) We want to utilize the abundant pixel-level prior information in the training data to construct a regularization term to constrain the model optimization, aiming to exclude erroneous mappings as much as possible.}
    \label{fig_1}
\end{figure}
While unsupervised cycle-consistent methods have demonstrated remarkable performance in various image-to-image translation tasks, they may produce multiple solutions (Figure \ref{fig_1}(a)), making them unsuitable for medical image-to-image translation tasks. 
Regularization methods\cite{b21,b22} incorporate prior constraints into the loss function to guide the model to choose gradient descent directions that satisfy these constraints during optimization, effectively narrowing the solution space of the model and improving the stability of its solutions.
Therefore, we assume that by extracting pixel-level prior information as much as possible from misaligned data and using it to develop a regularization term to constrain the training process, we should be able to effectively restrict the solution space of the unsupervised cycle-consistent GAN model.
This would enable the model to exclude erroneous mappings as much as possible (Figure \ref{fig_1}(b)), continuously approaching the unique optimal solution, thereby improving the performance and reliability of the generator.
To facilitate subsequent descriptions, we need to briefly analyze the misalignment errors in multi-modal medical images before introducing the proposed method.

\subsection{Registrable and unregistrable misalignment errors}
\begin{figure}[!htb]
  \centering  
  \includegraphics[width=6cm]{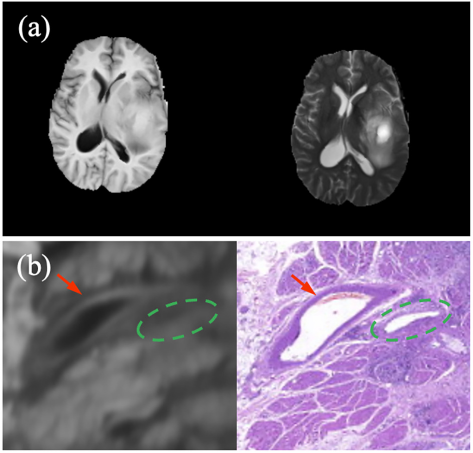}
  \caption{(a) T1 and T2 MR images of the same brain slice with affine deformation. (b) CT image and digital pathological image after H\&E staining of the same human cheek tissue sample.
  \label{fig_2}}  
\end{figure}
Medical images can be viewed as collections of anatomical structures.
Based on whether the misalignment errors in multi-modal medical image data can be repaired through registration, we can classify them into registrable misalignment errors and unregistrable misalignment errors.

\begin{figure*}[t]
    \centering  
    \includegraphics[width=16cm]{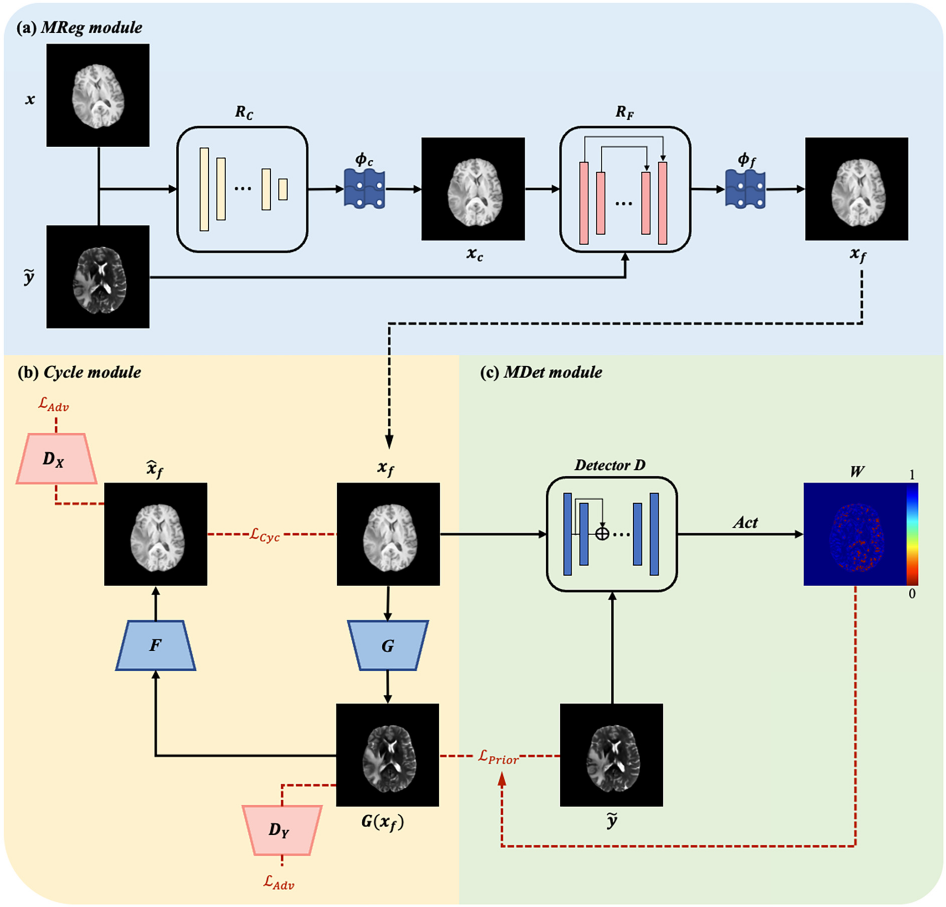}
    \caption{A general overview of MITIA. MITIA consists of three modules MDet, MReg, and Cycle. MReg is a multi-modal registration module. MDet is a multi-modal misalignment error detection module. Cycle is a cycle-consistent GAN-based image-to-image translation module.
    \label{fig_3}}  
\end{figure*}

\textbf{Registrable misalignment errors} are mainly caused by affine deformation or slight elastic deformation, which do not affect the consistency of anatomical structures between different modal images.
Therefore, this type of misalignment error can be repaired by registration methods.
For example, as shown in Figure \ref{fig_2}(a), T1 and T2 MR images of the same brain slice exhibit only misalignment errors caused by affine deformation. 
Assuming that the T1 modality image $I_{t1}$ is composed of $n$ anatomical structures $F_{t1}=\left\{f_{t1}^1,f_{t1}^2,...,f_{t1}^n\right\}$, and since affine deformation only changes the spatial position of each anatomical structure without causing any structural loss, the T2 modality image $I_{t2}$ must also be composed of $n$ anatomical structures $F_{t2}=\left\{f_{t2}^1,f_{t2}^2,...,f_{t2}^n\right\}$.
The anatomical structures in $F_{t1}$ and $F_{t2}$ can be one-to-one correspondence. In this case, a deformation field $\phi$ can be found such that $I_{t1}\circ\phi$ is pixel-wise aligned with $I_{t2}$ (where $\circ$ denotes the resampling operation), thereby correcting the misalignment errors between $I_{t1}$ and $I_{t2}$.

\textbf{Unregistrable misalignment errors} are mainly caused by anatomical structure loss due to sample variations between different modal imaging, which cannot be corrected by registration.
For example, as shown in Figure \ref{fig_2}(b), CT image $I_{ct}$ and digital pathological image after H\&E staining $I_{wsi}$ of the same human cheek tissue sample exhibit misalignment errors caused by anatomical structure loss due to tissue overlap (red arrows) and tissue tearing (green area) in $I_{wsi}$. 
In this case, where the anatomical structures are not completely consistent between $I_{ct}$ and $I_{wsi}$, assuming $I_{ct}$ is composed of n anatomical structures $F_{ct}=\left\{f_{ct}^1,f_{ct}^2,...,f_{ct}^n\right\}$, there must exist $f_{ct}^i\in F_{ct}\left(1\le i\le n\right)$ which cannot find the correspondence in $F_{wsi}=\left\{f_{wsi}^1,f_{wsi}^2,...,f_{wsi}^m\right\}$. 
Therefore, no registration method can repair the missing anatomical structures in $I_{wsi}$, and we refer to misalignment errors caused by anatomical structure loss as unregistrable misalignment errors.

\subsection{MITIA}
MITIA consists of three modules MDet, MReg and Cycle, as shown in Figure \ref{fig_3}.
We begin by formulating our MITIA model along the way introducing our notation.
Suppose $\{(x_i,{\widetilde{y}}_i)\}_{i=1}^n$ represents the dataset of misaligned multi-modal medical images, where $x_i$ and ${\widetilde{y}}_i$ come from modalities $X$ and $Y$ respectively, and ``\textasciitilde'' indicates the presence of misalignment errors between them.
Let $y_i$ be a modality $Y$ image that is pixel-wise aligned with $x_i$, but only exists theoretically.
The aim of this paper can be described as training a reliable ``$X\rightarrow Y$'' generator $G$ under the condition of only having the misaligned dataset $\{(x_i,{\widetilde{y}}_i)\}_{i=1}^n$.

For each pair of misaligned images $(x,\widetilde{y})$, we first extract pixel-level prior information from them. 
Then, we use the extracted prior information to develop a regularization term to constrain the training process. 
Since there are misalignment errors in $(x,\widetilde{y})$, only the regions in $\widetilde{y}$ where the anatomical structures are already aligned with $x$ can provide correct prior information for optimizing the model, while other regions provide incorrect prior information. 
If all the pixel-level prior information in $\widetilde{y}$ is used to constrain the model optimization, the corresponding prior regularization term $Q(x,\widetilde{y})$ can be described as follows:
\begin{equation}
	\label{eq_1}
  \begin{aligned}
    Q(x,\widetilde{y}) &= \mathbb{E}_{x,\widetilde{y}}\left[ || G(x) - \widetilde{y} ||_1 \right] \\
    &= \mathbb{E}_{x,\widetilde{y}}\left[ || (G(x) - \widetilde{y}) \cdot M_{\Omega} + (G(x) - \widetilde{y}) \cdot M_{\bar{\Omega}}||_1 \right],
  \end{aligned}
\end{equation}
where $\Omega$ represents the regions of $\widetilde{y}$ containing correct prior information, $\bar{\Omega}$ represents the rest regions containing incorrect prior information, $M_{\Omega}$ and $M_{\bar{\Omega}}$ are masks for $\Omega$ and $\bar{\Omega}$, respectively. 
It can be seen that when the correct prior information in $\Omega$ guides the model to optimize in the correct direction, the incorrect prior information in $\bar{\Omega}$ will mislead the model to optimize in other wrong directions.
This contradiction will prevent the prior regularization term $Q(x,\widetilde{y})$ from playing its proper role, leading to an unstable training process and ineffective improvement in the performance of the generator $G$.
Therefore, eliminating the interference of incorrect prior information in $\bar{\Omega}$ on the training process is crucial to ensure the intended function of the prior regularization term $Q(x,\widetilde{y})$.
Due to the complexity of the pixel-level prior information in $\widetilde{y}$, it is difficult to manually annotate the correct prior information. 
Thus, we use a deep neural network with powerful feature extraction capability to extract the correct prior information.
This prior extraction network consists of two pre-trained modules, the multi-modal registration module MReg and the multi-modal misalignment error detection module MDet.

\subsubsection{MReg Module}
MReg (Figure \ref{fig_3}(a)) aims to eliminate registrable misalignment errors in $(x,\widetilde{y})$, enabling $\widetilde{y}$ to provide more correct pixel-level prior information. 
To achieve better registration results, we adopt a coarse-to-fine cascaded registration method. 
The coarse registration model $R_C$ is trained under the constraint of mutual information loss\cite{b23,b24} $\mathcal{L}_{Coarse}$ (Equation \ref{eq_2}) to learn an affine deformation field $\phi_c=R_C(x,\widetilde{y})$, maximizing the mutual information between $x\circ\phi_c$ and $\widetilde{y}$.
\begin{equation}
  \label{eq_2}
  \mathcal{L}_{Coarse} = -\sum_{i,j}{P_{x,\widetilde{y}}\left(i,j\right)log{\left(\frac{P_{x,\widetilde{y}}\left(i,j\right)}{P_{x}\left(i\right)P_{\widetilde{y}}\left(j\right)}\right)}}
\end{equation}
Here, $P_{x,\widetilde{y}}\left(i,j\right)$ represents the joint probability of pixel values $(i, j)$ in the two images. $P_x\left(i\right)$ and $P_{\widetilde{y}}\left(j\right)$ represent the marginal probability distributions of pixel values $i$ and $j$ in images $x$ and $\widetilde{y}$, respectively.
The coarse registration model can globally correct misalignment errors caused by substantial yet relatively regular affine deformations, thereby reducing the workload of the fine registration model.
The fine registration model $R_F$ optimizes its parameters by minimizing the error output of the multi-modal misalignment error detector $D$ (to be detailed in Section II.D) in the pretrained MDet module, as shown in Equation (\ref{eq_3}).
This optimization allows $R_F$ to generate a more accurate deformation field $\phi_f=R_F(x\circ\phi_c,\widetilde{y})$, correcting the slight but irregular elastic deformation errors present in the image pair $(x\circ\phi_c,\widetilde{y})$ obtained after coarse registration. Additionally, to ensure that $R_F$ generates a smooth deformation field, we introduce an additional diffusion regularizer\cite{b25} on the gradient of the deformation vector field to constrain $\phi_f$ (Equation \ref{eq_4}). The overall objective of the fine registration model $R_F$ can be represented as Equation (\ref{eq_5}):
\begin{equation}
  \label{eq_3}
  \mathcal{L}_{Error} = \mathbb{E}_{x,\widetilde{y}}[||D(x \circ R_C(x,\widetilde{y}),\widetilde{y})||_1] 
\end{equation}
\begin{equation}
  \label{eq_4}
  \mathcal{L}_{Smooth} = \mathbb{E}_{x,\widetilde{y}}[|| \nabla R_F (x \circ R_C(x, \widetilde{y}), \widetilde{y}) ||^2]      
\end{equation}
\begin{equation}
  \label{eq_5}
  \mathcal{L}_{Fine} = \mathcal{L}_{Error}+{\lambda_{Smooth}\mathcal{L}}_{Smooth}
\end{equation}
Finally, the MReg module will produce a complete deformation field $\phi=\phi_c+\phi_f$. 
By applying $\phi$ to $x$, a new image pair $(x\circ\phi,\widetilde{y}) = (x_f, \widetilde{y})$ is generated to rectify the registrable misalignment errors in $(x,\widetilde{y})$.
If we denote the regions in $\bar{\Omega}$ of $\widetilde{y}$ where registrable misalignment errors exist as ${\bar{\Omega}}_R$, and the regions where unregistrable misalignment errors exist as ${\bar{\Omega}}_I$, according to our analysis of misalignment errors in Section II.B, we have $\bar{\Omega}={\bar{\Omega}}_R+{\bar{\Omega}}_I$. 
Thus, the corresponding prior regularization term $Q(x,\widetilde{y})$ after incorporating the MReg module can be described as follows:
\begin{equation}
	\label{eq_6}
  \begin{aligned}
    Q(x,\widetilde{y}) &= \mathbb{E}_{x,\widetilde{y}}\left[ || G(x_f) - \widetilde{y} ||_1 \right] \\
    &= \mathbb{E}_{x,\widetilde{y}}\left[ || (G(x_f) - \widetilde{y}) \cdot (M_{\Omega+\bar{\Omega}_R} + M_{\bar{\Omega}_I})||_1 \right]
  \end{aligned}
\end{equation}
It is evident that as the registration progresses, the regions of $\widetilde{y}$ containing correct prior information expands from the original $\Omega$ to $\Omega+{\bar{\Omega}}_R$, while the regions containing incorrect prior information shrinks from ${\bar{\Omega}}_R+{\bar{\Omega}}_I$ to ${\bar{\Omega}}_I$. 
Therefore, the introduction of MReg can increase correct pixel-level prior information to boost the training process.

\begin{figure*}[t]
    \centering  
    \includegraphics[width=16cm]{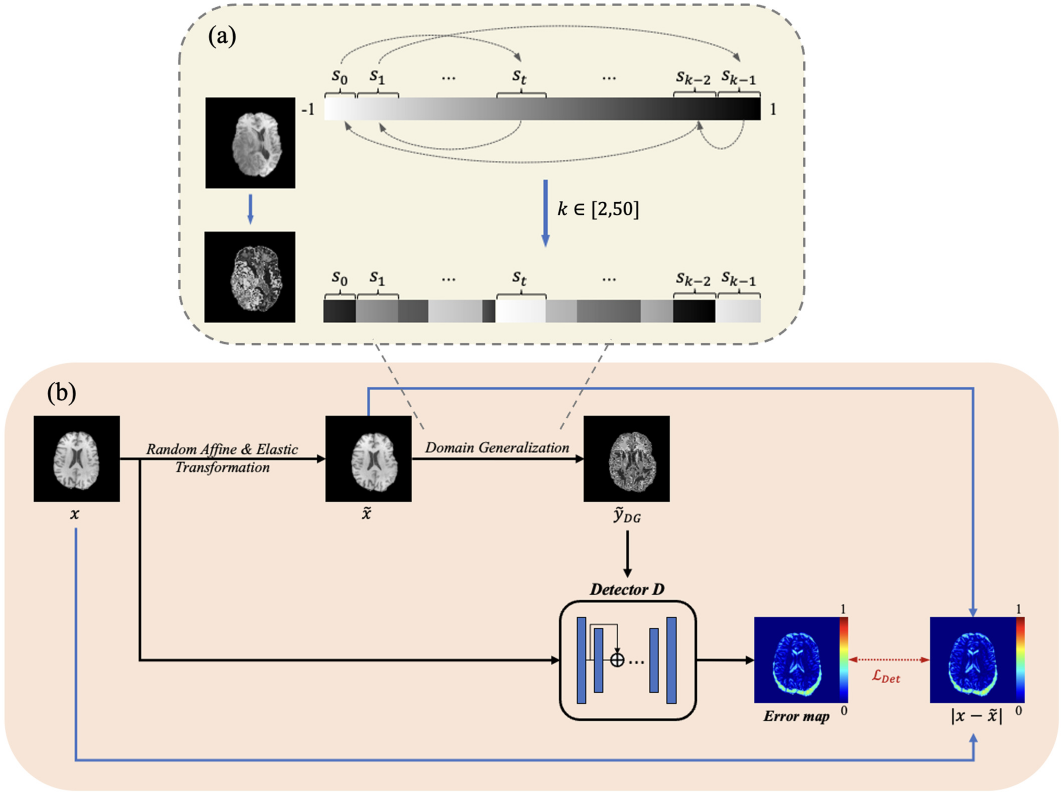}
    \caption{(a) Domain generalization method for simulating multiple different modalities. (b) Training process of the multi-modal misalignment error detector $D$.
    \label{fig_4}}  
\end{figure*}

\subsubsection{MDet Module}
Because the unregistrable misalignment errors in ${\bar{\Omega}}_I$ cannot be corrected by registration, some incorrect prior information contained in ${\bar{\Omega}}_I$ will still interfere with the model optimization. 
Therefore, we introduce MDet, as shown in Figure \ref{fig_3}(c).
MDet uses a multi-modal misalignment error detector $D$ to detect the remaining unregistrable misalignment errors in $(x_f,\widetilde{y})$, and uses an activation function $Act$ to activate the detection results, generating a confidence matrix $W=Act(D(x_f,\widetilde{y}))$ with the same dimension as the input image and a value range of $[0,1]$.
For regions identified by $D$ as having significant misalignment errors, MDet assigns very low weight values to the corresponding areas in the confidence matrix $W$, avoiding the incorporation of erroneous prior information in these regions into the prior regularization term.
For regions not identified as having significant misalignment errors, MDet assigns relatively high weight values to the corresponding areas in the $W$ matrix.
After introducing the MDet module, the prior regularization term $Q(x,\widetilde{y})$ in Equation (\ref{eq_6}) can be modified as follows:
\begin{equation}
	\label{eq_7}
  \begin{aligned}
    Q(x,\widetilde{y}) &= \mathbb{E}_{x,\widetilde{y}}\left[ || (G(x_f) - \widetilde{y}) \cdot W ||_1 \right] \\
    &= \mathbb{E}_{x,\widetilde{y}}\left[ || (G(x_f) - \widetilde{y}) \cdot (M_{\Omega+\bar{\Omega}_R}\cdot W + M_{\bar{\Omega}_I}\cdot W)||_1 \right]
  \end{aligned}
\end{equation}
Ideally, the regions in $W$ with a value of 0 should correspond to ${\bar{\Omega}}_I$, while the regions with a value of 1 should correspond to $\Omega+{\bar{\Omega}}_R$. 
Thus, the prior regularization term $Q(x,\widetilde{y})$ can be described as follows:
\begin{equation}
	\label{eq_8}
  \begin{aligned}
    Q(x,\widetilde{y}) &= \mathbb{E}_{x,\widetilde{y}}\left[ || (G(x_f) - \widetilde{y}) \cdot (M_{\Omega+\bar{\Omega}_R}\cdot 1 + M_{\bar{\Omega}_I}\cdot 0)||_1 \right] \\
    &= \mathbb{E}_{x,\widetilde{y}}\left[ || (G(x_f) - \widetilde{y}) \cdot M_{\Omega+\bar{\Omega}_R}||_1 \right]
  \end{aligned}
\end{equation}
At this point, the prior extraction network composed of the MReg and MDet modules has the ability to extract correct prior information and eliminate incorrect prior information from misaligned image pairs $\{(x_i,{\widetilde{y}}_i)\}_{i=1}^n$ as much as possible.

\subsubsection{Cycle Module}
After obtaining the pre-trained modules MReg and MDet, we can incorporate the following prior regularization loss $\mathcal{L}_{Prior}$,
\begin{equation}
  \label{eq_9}
  \mathcal{L}_{Prior} = \mathbb{E}_{x,\widetilde{y}}[|| (G(x_f) - \widetilde{y}) \cdot W ||_1],
\end{equation}
into the Cycle module, where $x_f=x\circ\phi=x\circ(R_C(x,\widetilde{y})+R_F(x\circ R_C(x,\widetilde{y}),\widetilde{y}))$ and $W=Act(D(x_f,\widetilde{y}))$.
Up to this point, the full objective of MITIA can be written as follows:
\begin{equation}
  \label{eq_10}
  \mathcal{L}_{Total} = \mathcal{L}_{Adv}+{\lambda_{Cyc}\mathcal{L}}_{Cyc}+{\lambda_{Prior}\mathcal{L}}_{Prior},
\end{equation}
in which $\mathcal{L}_{Cyc}$ is the cycle-consistency loss (Equation \ref{eq_11}) and $\mathcal{L}_{Adv}$ is the adversarial loss (Equation \ref{eq_12}).
\begin{equation}
  \label{eq_11}
  \mathcal{L}_{Cyc} = \mathbb{E}_{x_f} \left[ || F\left( G(x_f) \right) - x_f ||_1 \right] + \mathbb{E}_{\widetilde{y}} \left[ || G\left( F(\widetilde{y}) \right) - \widetilde{y} ||_1 \right] 
\end{equation}
\begin{equation}
	\label{eq_12}
  \begin{aligned}
    \mathcal{L}_{Adv} &= \mathbb{E}_{\widetilde{y}} \left[ log(D_Y(\widetilde{y})) \right] + \mathbb{E}_{x_f} \left[ log(1 - D_Y( G(x_f) )) \right] \\
                      &+ \mathbb{E}_{x_f} \left[ log(D_X(x_f)) \right] + \mathbb{E}_{\widetilde{y}} \left[ log(1 - D_X( F(\widetilde{y}) )) \right]
  \end{aligned}
\end{equation}
Here, $G$ and $F$ are generators, and $D_X$ and $D_Y$ are discriminators.

\subsection{Multi-modal misalignment error detector}
To effectively detect the remaining unregistrable misalignment errors in the image pair $(x_f,\widetilde{y})$ processed by MReg, inspired by the multi-modal spatial evaluator IMSE\cite{b26}, we adopt a training process as shown in Figure \ref{fig_4}(b) to train the detector $D$.
Firstly, we apply random affine and elastic deformations to the input image $x$ from modality $X$ to obtain a transformed image $\widetilde{x}$ that introduces signle-modal misalignment errors with respect to $x$.
Then, we employed a domain generalization method called Shuffle Remap\cite{b26} (as shown in Figure \ref{fig_4}(a)). Specifically, this method randomly divides the distribution of $\tilde{x}$ into $k$ segments, where $k \in [2,50]$ is a random number, then shuffles these segments and remaps them in the shuffled order to simulate the distribution of images from different modalities.
Hence, we can obtain an image ${\widetilde{y}}_{DG}$ that is pixel-wise aligned with $\widetilde{x}$ but exhibits multi-modal misalignment errors with $x$. 
Directly quantifying the multi-modal misalignment errors between $x$ and ${\widetilde{y}}_{DG}$ is difficult, but the single-modal misalignment errors between $x$ and $\widetilde{x}$ can be easily quantified using the residual map $\left|x-\widetilde{x}\right|$ between them. 
Therefore, we consider normalizing the result of $\left|x-\widetilde{x}\right|$ as the training label to train detector $D$ to convert the multi-modal misalignment errors between $x$ and ${\widetilde{y}}_{DG}$ into the single-modal misalignment errors between $x$ and $\widetilde{x}$.
Through this training, detector $D$ can quantify the multi-modal misalignment errors and provide an error map ranging from $[0,1]$. The optimization objective of detector $D$ can be represented as Equation (\ref{eq_13}).
\begin{equation}
  \label{eq_13}
  \mathcal{L}_{Det} = \mathbb{E}_{x} \left[|| D(x, \widetilde{y}_{DG}) - | x - \widetilde{x} | ||_1\right]
\end{equation}

\begin{figure}[!htb]
  \centering  
  \includegraphics[width=6cm]{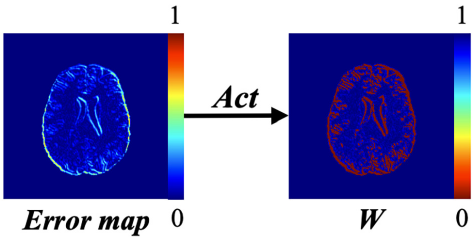}
  \caption{Convert the error map output by $D$ into a confidence matrix $W$ using the activation function $Act$.
  \label{fig_5}}  
\end{figure}

However, the output of $D$ cannot be directly used as the confidence matrix $W$, so we still need to do some post-processing on it (Figure \ref{fig_5}).
When using $D$ to detect the image pair $(x_f,\widetilde{y})$ processed by MReg, if a region of the error map output by $D$ has an error value greater than a threshold $th$, it can be determined that there is a significant misalignment error between this region of $x_f$ and $\widetilde{y}$.
Therefore, we can directly set the weight value of the corresponding region of $W$ to 0. Since the magnitude of the single-modal residual value is also related to the pixel values of the image itself, for regions with residual values less than $th$, their weights still need to be appropriately reduced based on the magnitude of their residuals.
Considering the above factors, the final confidence matrix $W$ and activation function $Act$ are as shown in Equation (\ref{eq_14}):
\begin{equation}
  \label{eq_14}
  \begin{aligned}
    W &= Act(D(x_f, \widetilde{y})) \\
      &= 1 - D(x_f, \widetilde{y}) \cdot M_{th}
  \end{aligned}
\end{equation}
where $M_{th}$ represents the mask for regions where the residual value is less than $th$.

\section{Experiments and results}
\begin{figure*}[!htb]
    \centering  
    \includegraphics[width=16cm]{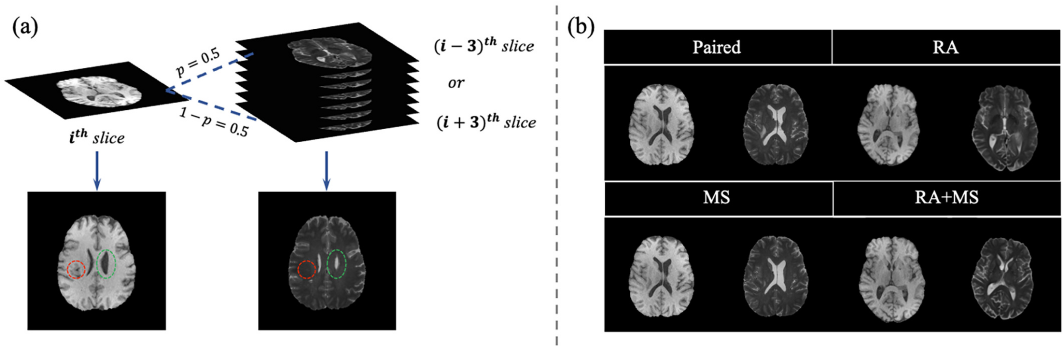}
    \caption{(a) Constructing misaligned image pairs using the Mis-Slice method. (b) Four modes for constructing training sets that introduce different misalignment errors.
    \label{fig_6}}  
\end{figure*}
\subsection{Datasets}
We evaluated MITIA using two publicly available datasets, BraTS2020\cite{b27} and PDGM\cite{b28}, where the original images of different modalities are well-aligned. 
We introduced misalignment errors using two methods, Random-Affine and Mis-Slice, to create training sets of multi-modal medical images with different types and severity of misalignment errors.
The \textbf{Random-Affine} method introduces registrable misalignment errors caused by affine deformation by randomly adding $[-3, +3]$ degrees of rotation, $[-3\%, +3\%]$ of translation, and $[-3\%, +3\%]$ of scaling to the original images.
The \textbf{Mis-Slice} method (Figure \ref{fig_6}(a)) introduces unregistrable misalignment errors caused by the absence of anatomical structures (red region in Figure \ref{fig_6}(a)) and registrable misalignment errors caused by elastic deformation (green region in Figure \ref{fig_6}(a)) by randomly pairing the $i^{th}$ slice of the volume data from modality $X$ with the $(i\pm3)^{th}$ slice of the volume data from modality $Y$ with a probability of $p=0.5$.
We designed four training set construction modes that can introduce different misalignment errors as follows (Figure \ref{fig_6}(b)):
\begin{itemize}
  \item \textbf{Paired}: Construct training sets using well-aligned original images, and it does not introduce misalignment errors.
  \item \textbf{RA}: Construct training sets using Random-Affine method, and it introduces only registrable misalignment errors.
  \item \textbf{MS}: Construct training sets using Mis-Slice method, and it introduces unregistrable misalignment errors and a small amount of registrable misalignment errors.
  \item \textbf{RA+MS}: Construct training sets using both Random-Affine method and Mis-Slice method, and it introduces unregistrable misalignment errors and significant registrable misalignment errors.
\end{itemize}
In BraTS2020, we selected 240 pairs of T1-T2 volumes, and in PDGM, we selected 160 pairs of T2-FLAIR volumes.
For each pair of volumes, we selected 50 pairs of axial cross-sections with brain tissue to construct four training sets using Paired, RA, MS, and RA+MS modes.
In the end, all training sets constructed by BraTS2020 contain 12000 image pairs, while all training sets constructed by PDGM contain 8000 image pairs.
Additionally, we randomly selected 1000 paired T1-T2 images from BraTS2020 and 800 paired T2-FLIAR images from PDGM to construct two test sets.
All images were standardized to the range $[-1,1]$ and resampled to a size of $256\times256$.
Finally, we obtained six misaligned training sets with different misalignment errors, two well-aligned training sets, and two well-aligned test sets.

\subsection{Implementation and training details}
In the MReg module, the coarse registration model $R_C$ consists of five $3\times3$ convolutional layers and two fully connected layers.
The second and fourth convolutional layers are followed by $2\times2$ max pooling operations with a stride of 2.
The stride of the first convolutional operation is 2, while the strides of the remaining convolutional operations are 1. 
Each convolutional operation is followed by Batch Normalization\cite{b29} and Leaky-ReLU activation. 
The number of filters for the five convolutional layers and the final two fully connected layers is set as follows: $[32, 64, 64, 64, 64, 32, 4]$.
Ultimately, $R_C$ outputs a set of parameters $\theta_c$ representing rotation, scaling, and translation, from which we can obtain the affine deformation field $\phi_c$. 
The fine registration model $R_F$ is based on UNet\cite{b30}. The number of filters for the downsampling layers is set as follows: $[32, 64, 64, 64, 64, 64, 64]$, and the number of filters for the upsampling layers is set as follows: $[64, 64, 64, 64, 64, 64, 32]$.
After the upsampling layers, $R_F$ directly outputs a deformation field $\phi_f$ with 2 channels through a $3\times3$ convolutional layer.
The multi-modal misalignment error detector $D$ in the MDet module and the generators $G$ and $F$ in the Cycle module are consistent with the generator in CycleGAN, which contains 9 res-blocks\cite{b14,b31}.
The discriminators $D_X$ and $D_Y$ in the Cycle module are based on PatchGAN\cite{b13}. The activation threshold $th$ for the activation function $Act$ in MDet is set to $0.1$.
The network was implemented based on the PyTorch framework and was performed on a computer with an Nvidia GeForce RTX 4090 GPU. 
The batch size was set to 1, and the training epochs for both the MReg and MDet modules were set to 80, while the Cycle module was trained for 60 epochs.
All loss functions were optimized using the Adam optimizer with $(\beta_1,\beta_2)=(0.5, 0.999)$ and a learning rate of $1e-4$. 
The weights for the loss functions were set to $\lambda_{Smooth}=1$, $\lambda_{Cyc}=10$, and $\lambda_{Prior}=30$.

\subsection{Competing methods}
We compared MITIA with several state-of-the-art image-to-image translation methods, including supervised GAN (Pix2Pix\cite{b13}, RegGAN\cite{b6}), unsupervised GAN (CycleGAN\cite{b14}, UNIT\cite{b15}, MUNIT\cite{b16}), and diffusion models (SynDiff\cite{b7}).
Pix2Pix is a typical supervised GAN consisting of a generator $G$ and a discriminator $D$, optimizing the generator by minimizing the pixel-wise loss between the predicted image $G(x)$ and the target image $y$.
Pix2Pix performs well when training data is highly aligned.
RegGAN, based on the ``loss-correction'' theory, extends Pix2Pix by introducing a registration network to fit the misalignment noise distribution between the predicted image $G(x)$ and the target image $y$, enabling better performance in the presence of misalignment errors introduced by affine or elastic deformation in the training data.
Since MITIA uses a network structure with two generators and two discriminators, we implemented two additional comparative methods based on Pix2Pix and RegGAN with a similar structure to ensure consistency in network structure for a fair comparison of different methods' performance.
These two methods introduce an additional generator and discriminator to both Pix2Pix and RegGAN, and incorporate a cycle-consistency loss with a weight $\lambda_{Cyc}=10$ into their original objective functions.
The weights of the other loss terms in the original objective functions were kept unchanged, with the weight of the pixel-wise loss constraint $\mathcal{L}_{L1}$ in Pix2Pix being $\lambda_{L1}=100$ and the weight of the correction loss constraint $\mathcal{L}_{Corr}$ in RegGAN being $\lambda_{Corr=20}$.
Since these two additional comparative methods have not been proposed in previous work, we refer to them as Cyc-Pix2Pix and Cyc-RegGAN, respectively.
CycleGAN is the most representative unsupervised cycle-consistent GAN, which completes the inverse mapping of $G:X\rightarrow Y$ by adding a reverse generator $F:Y\rightarrow X$, and introduces cycle-consistency loss to enforce $F(G(X))\approx X$ and $G(F(Y))\approx Y$, thus enabling training of the model without paired data.
The variant of CycleGAN, UNIT, assumes that the source domain and the target domain share a latent space, mapping the source domain image $x$ and the target domain image $y$ to the same latent code to establish the relationship between the two domains. 
MUNIT further assumes a shared content space based on UNIT, completing the translation task by decoupling and recombining image content and style information.
The unsupervised SynDiff is the latest attempt of diffusion models in the field of multi-modal medical image-to-image translation. It implements fine image sampling through conditional diffusion processes to capture the correlation between the distributions of images from different modalities, while introducing cycle-consistency loss and discriminator loss to enable training on unpaired datasets.
Compared to previous registration and image-to-image translation methods, such as RegGAN and Cyc-RegGAN, the proposed MITIA method not only employs a more effective coarse-to-fine registration module, MReg, which is independently trained under registration loss to provide more available pixel-level prior information for model optimization, but also incorporates an error detection module, MDet, to prevent unregistrable misalignment errors from interfering with model training. With the aid of these two modules, MITIA can maximize the use of pixel-level prior information available in the training data to guide model optimization, thereby effectively enhancing the performance of the generator.

\subsection{Results and analysis}
\subsubsection{Demonstration of the prior extraction network's role in addressing misalignment errors}
\begin{figure}[!htb]
  \centering  
  \includegraphics[width=8cm]{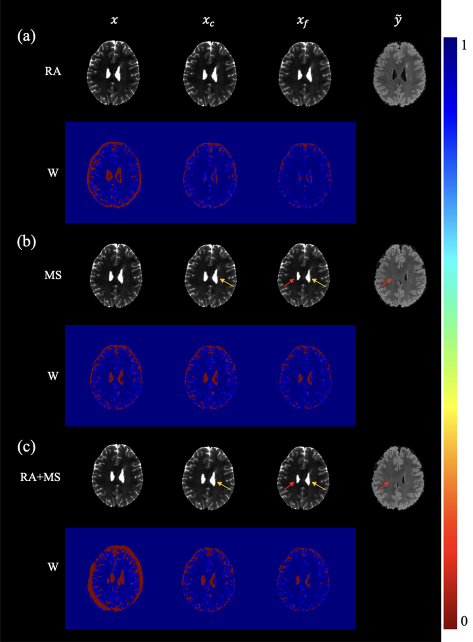}
  \caption{The image results through $R_C$ and $R_F$, along with the corresponding confidence matrices $W$ on the RA, MS, and RA+MS training sets constructed by PDGM.}
  \label{fig_7}  
\end{figure}
To intuitively demonstrate the roles of each component of the prior extraction network in addressing different types of misalignment errors, we utilized the pretrained MReg module to perform coarse-to-fine registration on training data with different types of misalignment errors and utilized the pretrained MDet module to detect the misalignment errors.
The experiments in this section were conducted on the RA, MS, and RA+MS training sets constructed by T2-FLAIR volume data from PDGM.
Table \ref{table_1} lists the average misalignment errors obtained from detector D based on the training sets.
Figure \ref{fig_7} presents the image results through $R_C$ and $R_F$, along with the corresponding confidence matrices $W$.
The first column of Table \ref{table_1} and Figure \ref{fig_7} shows the results before the registration, while the second and third columns show the results after coarse registration $R_C$ and fine registration $R_F$, respectively.
As shown in the confidence matrices $W$ in Figure \ref{fig_7} and the quantitative results in Table \ref{table_1}, the misalignment errors between $x$ and $\tilde{y}$ are notably reduced after $R_C$ on RA and RA+MS.
In contrast, the misalignment errors between $x$ and $\tilde{y}$ show unnoticeable change before and after $R_C$ on MS.
\begin{table}[t]
  \begin{center}
    \caption{The average misalignment errors through $R_C$ and $R_F$ on RA, MS, and RA+MS training sets constructed by PDGM.}
    \label{table_1}
    \begin{tabular}{cccc}
    \toprule[1.5pt]
            &  \multicolumn{3}{c}{Average misalignment errors of training sets (\%)} \\
    \cmidrule{2-4}
            &      Before MReg    &      After $R_C$     &    After $R_F$  \\
    \midrule
    RA     &      2.78$\pm$0.87    &     1.16$\pm$0.30    &   0.93$\pm$0.23 \\
    MS      &     1.49$\pm$0.36    &     1.47$\pm$0.36    &   1.12$\pm$0.29 \\
    RA+MS   &     2.83$\pm$0.89    &     1.50$\pm$0.37    &   1.14$\pm$0.29 \\
    \bottomrule[1.5pt]
    \end{tabular}
  \end{center}
\end{table}
This indicates that $R_C$ can effectively reduce registerable misalignment errors caused by affine deformation but has negligible effect on misalignment errors caused by elastic deformation or missing anatomical structures.
After $R_F$, the average error values on MS and RA+MS in Table \ref{table_1} show a noticeable reduction. As indicated by the yellow arrows in Figure \ref{fig_7}, $x_f$ is more structurally consistent with $\tilde{y}$ than $x_c$.
This demonstrates the effectiveness of $R_F$ in reducing misalignment errors caused by elastic deformation.
As shown by the red arrows in Figure \ref{fig_7}, there exists misalignment error between $x_f$ and $\tilde{y}$ due to missing anatomical structures.
These unregistrable misalignment errors are accurately detected by MDet and depicted in the confidence matrices $W$ in the third column of Figure \ref{fig_7}(b) and (c).
The above qualitative and quantitative results demonstrate that when addressing misaligned training data, the MReg module can effectively correct registrable misalignment errors, while the MDet module can accurately detect unregistrable misalignment errors.
The collaboration of these two modules increases the available pixel-level prior information in the training data while preventing unregistrable misalignment errors from interfering with model optimization.

\subsubsection{Performance on misaligned datasets}
\begin{figure*}[!htb]
  \centering  
  \includegraphics[width=16cm]{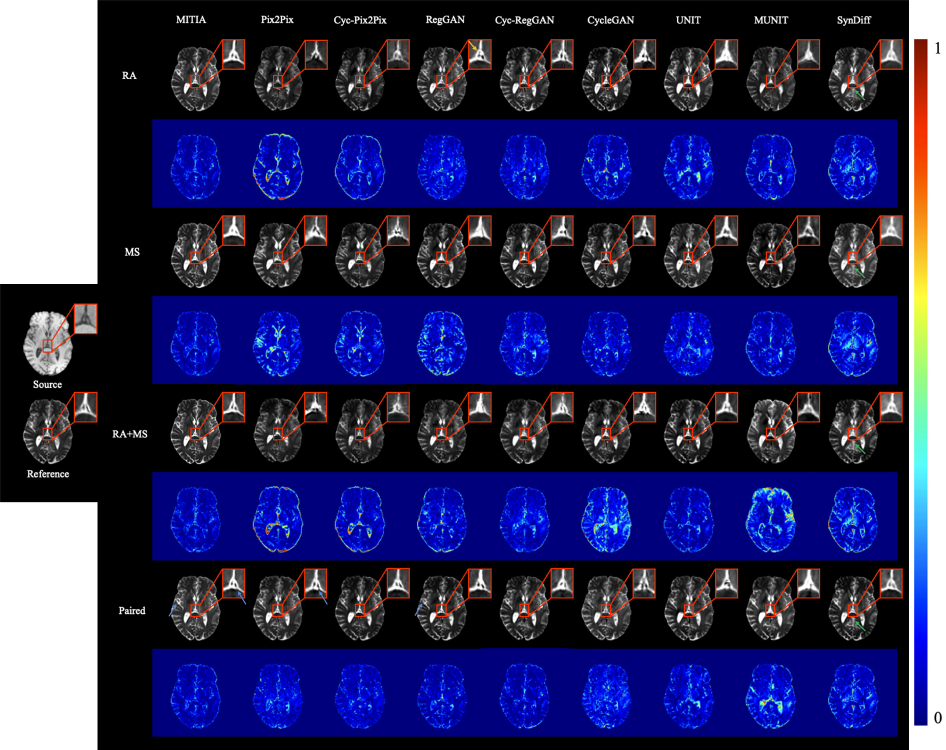}
  \caption{Qualitative comparison of different methods on the BraTS2020 dataset.
  \label{fig_8}}  
\end{figure*}
\begin{table*}[!htb]
    \begin{center}
      \caption{Comparison of PSNR and SSIM for different methods on the RA, MS, and RA+MS training sets constructed using BraTS2020.
      \label{table_2}
      }
      \begin{tabular}{ccccccc}
      \toprule[1.5pt]
                       &     \multicolumn{2}{c}{RA}    &     \multicolumn{2}{c}{MS}    &    \multicolumn{2}{c}{RA+MS}  \\
      \cmidrule{2-7}
                       &     PSNR(dB)  &    SSIM(\%)   &    PSNR(dB)   &    SSIM(\%)   &    PSNR(dB)   &    SSIM(\%)   \\
      \midrule
               Pix2Pix & 22.81$\pm$1.17 & 87.28$\pm$1.69 & 23.54$\pm$1.25 & 88.69$\pm$1.18 & 21.61$\pm$0.90 & 84.92$\pm$1.20 \\
           Cyc-Pix2Pix & 23.43$\pm$1.15 & 88.51$\pm$1.39 & 23.83$\pm$0.87 & 89.07$\pm$0.89 & 22.37$\pm$0.82 & 87.27$\pm$1.10 \\
               RegGAN  & 24.95$\pm$1.59 & 91.53$\pm$1.48 & 23.88$\pm$1.23 & 89.52$\pm$1.20 & 22.35$\pm$1.21 & 87.68$\pm$0.92 \\
           Cyc-RegGAN  & 24.56$\pm$1.56 & 91.09$\pm$1.32 & 24.38$\pm$1.29 & 90.12$\pm$1.35 & 24.15$\pm$1.43 & 89.83$\pm$1.31 \\
             CycleGAN  & 23.49$\pm$1.16 & 89.62$\pm$0.92 & 24.37$\pm$1.36 & 89.89$\pm$1.27 & 23.39$\pm$0.94 & 88.11$\pm$0.88 \\
                 UNIT  & 24.65$\pm$1.53 & 90.81$\pm$1.38 & 24.73$\pm$1.82 & 90.54$\pm$1.70 & 24.54$\pm$1.34 & 90.52$\pm$1.60 \\
                MUNIT  & 22.81$\pm$1.28 & 87.23$\pm$1.33 & 23.12$\pm$1.05 & 87.99$\pm$2.09 & 22.20$\pm$1.01 & 86.77$\pm$1.09 \\
              SynDiff  & 24.25$\pm$1.56 & 91.08$\pm$1.72 & 24.09$\pm$1.55 & 90.82$\pm$1.73 & 23.47$\pm$1.27 & 89.12$\pm$1.52 \\
          \textbf{MITIA(Ours)}  & \textbf{26.39$\pm$1.09} & \textbf{92.55$\pm$1.23} & \textbf{26.39$\pm$1.52} & \textbf{92.62$\pm$1.42} & \textbf{26.37$\pm$1.18} & \textbf{92.40$\pm$1.31} \\
      \bottomrule[1.5pt]
      \end{tabular}
    \end{center}
  \end{table*}

The purpose of designing MITIA is to enable reliable multi-modal medical image-to-image translation without relying on pixel-wise aligned data. 
Therefore, in this section, we trained MITIA using datasets containing different misalignment errors to demonstrate its feasibility and superiority. 
First, we trained seven different methods on the RA, MS, and RA+MS training sets constructed based on T1-T2 volume data from BraTS2020 and quantitatively evaluated the performance of all methods on the test set using peak signal-to-noise ratio (PSNR) and structural similarity (SSIM).
Quantitative results are listed in Table \ref{table_2}, while Figure \ref{fig_8} shows representative images and their corresponding residual maps compared to the reference image.
In the results of supervised GAN methods, the generated images of Pix2Pix exhibit noticeable errors in content, along with poor quantitative results, especially evident in RA+MS where its evaluation metrics are notably inferior to those in RA and MS.
It is expected because Pix2Pix heavily relies on well-aligned data and cannot avoid the interference of any misalignment errors in model optimization.
\begin{figure*}[!htb]
    \centering  
    \includegraphics[width=16cm]{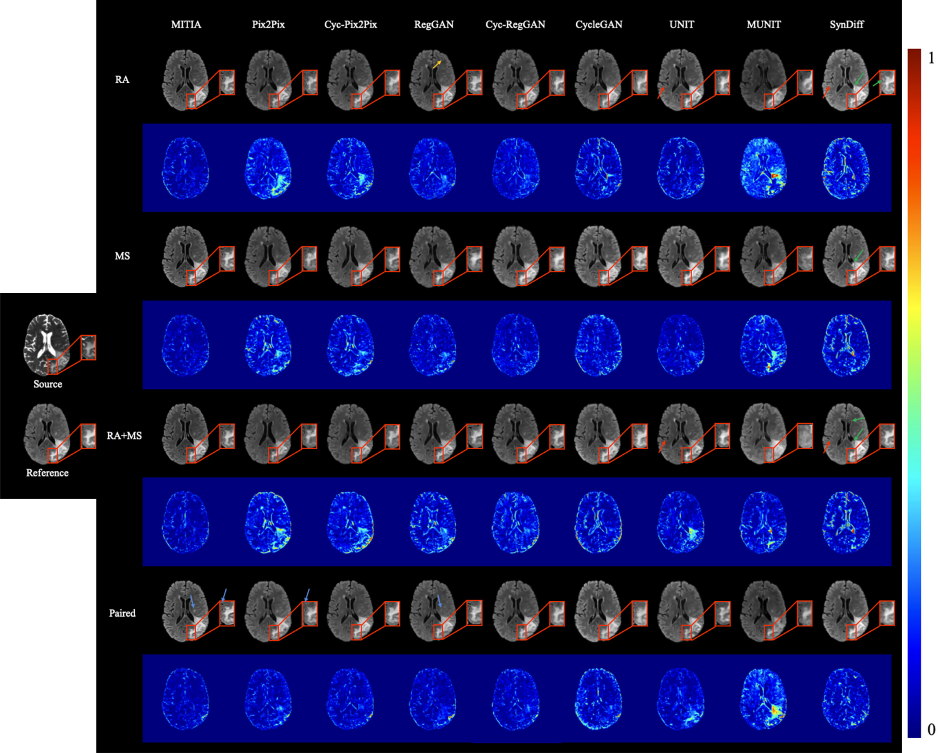}
    \caption{Qualitative comparison of different methods on the PDGM dataset.
    \label{fig_9}}  
  \end{figure*}  
\begin{table*}[!htb]
    \begin{center}
      \caption{Comparison of PSNR and SSIM for different methods on the RA, MS, and RA+MS training sets constructed using PDGM.
      \label{table_3}
      }
      \begin{tabular} {ccccccc}
      \toprule[1.5pt]
                       &     \multicolumn{2}{c}{RA}    &     \multicolumn{2}{c}{MS}    &    \multicolumn{2}{c}{RA+MS}  \\
      \cmidrule{2-7}
                       &     PSNR(dB)  &    SSIM(\%)   &    PSNR(dB)   &    SSIM(\%)   &    PSNR(dB)   &    SSIM(\%)   \\
      \midrule
               Pix2Pix & 24.32$\pm$0.93 & 87.72$\pm$1.06 & 23.75$\pm$0.86 & 85.14$\pm$1.34 & 21.94$\pm$0.69 & 84.65$\pm$1.30 \\
           Cyc-Pix2Pix & 24.63$\pm$1.09 & 88.65$\pm$0.96 & 23.80$\pm$0.97 & 86.96$\pm$0.85 & 23.39$\pm$0.74 & 85.75$\pm$1.35 \\
               RegGAN  & 25.82$\pm$1.36 & 91.56$\pm$0.88 & 24.96$\pm$1.29 & 88.28$\pm$0.79 & 23.54$\pm$0.87 & 85.21$\pm$1.43 \\
            Cyc-RegGAN & 25.80$\pm$1.38 & 90.80$\pm$0.86 & 25.36$\pm$1.33 & 89.28$\pm$0.78 & 24.61$\pm$0.95 & 87.52$\pm$1.56 \\
               CycleGAN  & 24.39$\pm$0.81 & 87.88$\pm$0.92 & 25.07$\pm$0.93 & 89.25$\pm$0.83 & 24.28$\pm$0.58 & 87.16$\pm$0.81 \\
                 UNIT  & 25.19$\pm$1.22 & 90.10$\pm$1.07 & 25.32$\pm$1.28 & 90.43$\pm$0.84 & 25.02$\pm$1.42 & 89.86$\pm$1.00 \\
                MUNIT  & 22.86$\pm$1.01 & 86.63$\pm$1.39 & 23.87$\pm$0.89 & 87.66$\pm$1.19 & 22.96$\pm$1.01 & 86.29$\pm$1.22 \\
              SynDiff  & 25.25$\pm$1.13 & 90.45$\pm$1.64 & 25.37$\pm$1.50 & 91.06$\pm$0.85 & 24.40$\pm$0.93 & 88.76$\pm$1.91 \\
          \textbf{MITIA(Ours)}  & \textbf{26.96$\pm$1.33} & \textbf{92.37$\pm$0.70} & \textbf{27.04$\pm$1.41} & \textbf{92.52$\pm$0.80} & \textbf{26.90$\pm$1.23} & \textbf{92.34$\pm$0.56} \\
      \bottomrule[1.5pt]
      \end{tabular}
    \end{center}
  \end{table*}
Thus, the performance of Pix2Pix deteriorates with an increasing presence of misalignment errors in the training data.
The performance metrics of Cyc-Pix2Pix are higher than those of Pix2Pix, especially showing greater advantages in RA+MS. From the residual maps, it is also visually evident that Cyc-Pix2Pix generates images with fewer errors compared to Pix2Pix. This indicates that compared to the network structure consisting of one generator and one discriminator, the network structure with two generators and two discriminators incorporating cycle-consistency constraints can mitigate the interference of misalignment errors during model training.
RegGAN achieves PSNR and SSIM scores second only to MITIA in RA, but its scores in MS and RA+MS are unsatisfactory.
Additionally, the image quality of RegGAN in MS and RA+MS does not show remarkable improvement compared to Pix2Pix.
This indicates that while the registration network in RegGAN can effectively mitigate the interference of registrable misalignment errors on model optimization, it cannot properly handle unregistrable misalignment errors.
Cyc-RegGAN achieves better quantitative results than RegGAN in both MS and RA+MS, and it depicts image details more accurately. However, Cyc-RegGAN performs worse than RegGAN in RA. The above results suggest that combining RegGAN with cycle-consistency constraints may be more effective in mitigating the interference of unregistrable misalignment errors during model training. However, when the training data contains only registrable misalignment errors, cycle-consistency constraints may play a negative role when combined with RegGAN, which is consistent with the conclusion in RegGAN\cite{b6}.
The quantitative results of unsupervised GAN methods are relatively insensitive to different types and severity of misalignment errors compared to supervised GAN methods. 
Among them, CycleGAN and UNIT generally outperform supervised GAN methods, except for being inferior to RegGAN in RA, and this superiority is most pronounced in RA+MS.
This is because they do not need pixel-level prior information in the training data to constrain model optimization.
However, the lack of guidance from pixel-level prior information also leads to these methods having poorer fidelity to image content, especially to some fine structures.
This deficiency is most evident in the results of MUNIT, which may be due to information loss in the process of decoupling and recombining content and style.
SynDiff excels at preserving the overall structure of some tissues in the image and performs well among unsupervised methods, thanks to the excellent performance in generating high-quality images of diffusion models.
However, SynDiff still cannot guarantee the correctness of image content.
The zoomed areas of the SynDiff result images in Figure \ref{fig_8} have lower contrast compared to other methods, and areas indicated by the green arrows unexpectedly generate false content similar to tissues that do not actually exist.
This indicates that even unsupervised methods based on powerful diffusion models still have an ambiguous solution space due to the lack of pixel-wise prior constraints, leading to unstable and unreliable translation results.
In terms of PSNR and SSIM, MITIA achieves the highest scores in RA, MS, and RA+MS. In terms of image quality, MITIA demonstrates superior fidelity to the content information of images compared to other methods.

To validate the performance of MITIA in different multi-modal medical image-to-image translation tasks, the same seven methods mentioned above were employed to train on RA, MS, and RA+MS training sets constructed by T2-FLAIR volume data from PDGM.
Qualitative and quantitative analyses of the trained models were conducted on the test set.
Table \ref{table_3} presents quantitative results, while representative images along with their residual maps compared to the reference image are displayed in Figure \ref{fig_9}.
Apart from RegGAN achieving decent quantitative results and high-quality result images in RA, the overall performance of supervised GAN methods is notably affected by misalignment errors in the training data, especially by unregistrable misalignment errors.
It is worth noting that the PSNR and SSIM of RegGAN are both lower than those of MITIA, which is consistent with the results in Table \ref{table_2}.
As shown in the regions indicated by the yellow arrows in Figure \ref{fig_8} and Figure \ref{fig_9}, the generated images of RegGAN in RA exhibit some loss of detail structures. 
We speculate that this is due to the limited performance of the registration network in RegGAN, which cannot correct all registrable misalignment errors, leading to remaining misalignment errors that still interfere with the model optimization to some extent.
In the results of unsupervised GAN methods, CycleGAN and MUNIT exhibit unstable generated images and serious loss of content information.
In contrast, UNIT shows better image quality, but it suffers from blurry organ boundaries (as shown in the zoomed areas of the UNIT result image in Figure \ref{fig_9}).
Quantitatively, similar to the results in BraTS2020, the evaluation metrics of the three unsupervised GAN methods are not outstanding, but their fluctuations when facing different misalignment errors are smaller compared to supervised methods.
The quantitative results of SynDiff surpass other unsupervised methods in RA and MS but are lower than UNIT in RA+MS, which is consistent with the visual results.
From the images generated by SynDiff, it can be seen that the overall structure of some tissues is well preserved (as shown in the zoomed areas of the SynDiff result images in Figure \ref{fig_9}), but it loses more fine structural details compared to UNIT (red arrows), and this loss is most pronounced in RA+MS. 
Additionally, similar to the results in BraTS2020, SynDiff generates a small amount of erroneous content that does not actually exist (green arrows).
Compared to other methods, MITIA can generate result images with more accurate content and more details without being affected by misalignment errors.
Quantitatively, MITIA also shows substantial advantages, with its PSNR and SSIM performance metrics remaining stable when faced with different types and severity of misalignment errors. 
The stability of MITIA's performance implies that the prior extraction network composed of the MReg and MDet modules can markedly eliminate the interference of misalignment errors on model optimization.
The superiority of MITIA's performance indicates that the pixel-level prior information extracted from misaligned training data can effectively constrain the training process, resulting in a substantial improvement in the performance of the generator.

\subsubsection{Performance on well-aligned datasets}
\begin{table}[!htb]
    \begin{center}
      \caption{Comparison of PSNR and SSIM for different methods on well-aligned training sets constructed using BraTS2020 and PDGM.
      \label{table_4}
      }
      \resizebox{\linewidth}{!}{
      \begin{tabular} {ccccc}
      \toprule[1.5pt]
                       &     \multicolumn{2}{c}{BraTS2020}   &    \multicolumn{2}{c}{PDGM}     \\
      \cmidrule{2-5}
                       &     PSNR(dB)  &    SSIM(\%)   &    PSNR(dB)   &    SSIM(\%)  \\
      \midrule
               Pix2Pix & 25.44$\pm$1.85 & 92.10$\pm$1.55 & 25.96$\pm$1.54 & 91.73$\pm$0.87 \\
           Cyc-Pix2Pix & 25.47$\pm$1.77 & 92.15$\pm$1.42 & 26.02$\pm$1.79 & 91.79$\pm$1.01 \\
               RegGAN  & 25.56$\pm$1.75 & 92.26$\pm$1.33 & 26.14$\pm$1.77 & 91.94$\pm$0.89 \\
            Cyc-RegGAN & 25.51$\pm$1.35 & 92.20$\pm$1.54 & 26.06$\pm$1.60 & 91.86$\pm$0.83 \\
             CycleGAN  & 24.73$\pm$1.63 & 89.97$\pm$1.36 & 25.28$\pm$0.73 & 89.68$\pm$0.70 \\
                 UNIT  & 24.81$\pm$1.45 & 90.82$\pm$1.83 & 25.39$\pm$1.58 & 90.78$\pm$1.11 \\
                MUNIT  & 23.61$\pm$1.71 & 88.45$\pm$2.17 & 23.95$\pm$1.56 & 88.04$\pm$2.44 \\
              SynDiff  & 25.02$\pm$1.71 & 91.41$\pm$1.86 & 25.88$\pm$1.54 & 91.23$\pm$0.80 \\
          \textbf{MITIA(Ours)}  & \textbf{26.49$\pm$1.40} & \textbf{92.69$\pm$1.38} & \textbf{27.09$\pm$1.45} & \textbf{92.53$\pm$0.51}  \\
      \bottomrule[1.5pt]
      \end{tabular}
      }
    \end{center}
  \end{table}
\begin{figure*}[!htb]
    \centering  
    \includegraphics[width=16cm]{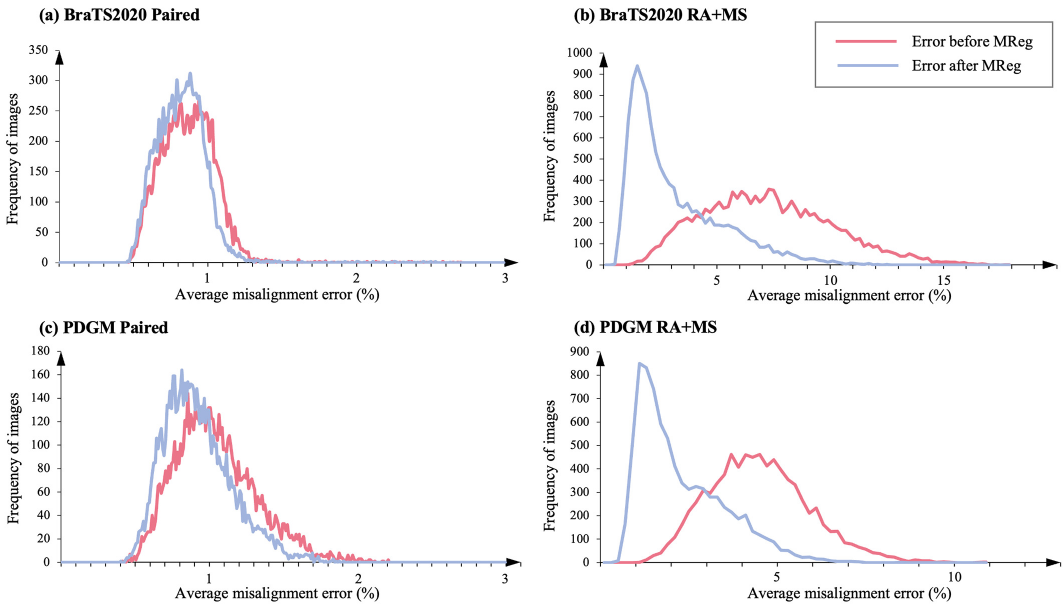}
    \caption{The red and blue lines are two frequency distribution line graphs used to represent the relationship between the average misalignment error per image and the frequency of images in the training set before and after processing by MReg.
    \label{fig_11}}  
  \end{figure*}
To comprehensively evaluate the performance of MITIA, in this section, we conducted further experiments on two well-aligned training sets constructed by T1-T2 volume data from BraTS2020 and T2-FLAIR volume data from PDGM, respectively. 
We quantitatively evaluated the trained models on the same two test sets as in Section III.D.1.
Quantitative results are presented in Table \ref{table_4}, and representative result images are shown in the fourth row of Figure \ref{fig_8} and Figure \ref{fig_9}.
When having highly aligned training sets, supervised methods show notable advantages in quantitative results and the quality of generated images compared to unsupervised methods, which once again demonstrates the importance of using pixel-level prior information to constrain model optimization for improving generator's performance and reliability.
With the substantial reduction of misalignment errors in the training data, the quantitative results of unsupervised methods also improve to varying degrees, indicating that highly aligned training data can reduce the difficulty of establishing relationships between different modalities for unsupervised methods.
However, from the result images, CycleGAN and MUNIT still have notable deficiencies. 
Although the image quality of UNIT is slightly improved, it still suffers from issues such as loss of detail structures and blurry organizational edges (as shown in the enlarged areas of UNIT results in Figure \ref{fig_8} and Figure \ref{fig_9}).
SynDiff, which performs best among unsupervised methods, still generates incorrect content information in the results as indicated by the green arrow in the fourth row of Figure \ref{fig_8}.
\begin{figure}[!htb]
  \centering  
  \includegraphics[width=8cm]{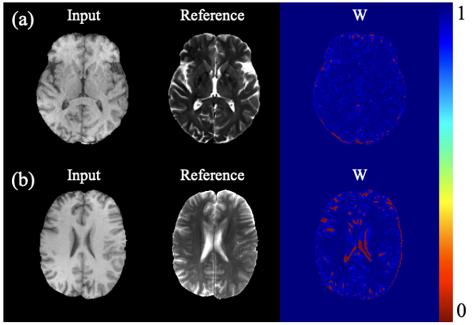}
  \caption{Aligned multi-modal image pairs and the corresponding confidence matrix $W$ output by MDet
  \label{fig_10}}  
\end{figure}

\begin{figure*}[!htb]
    \centering  
    \includegraphics[width=16cm]{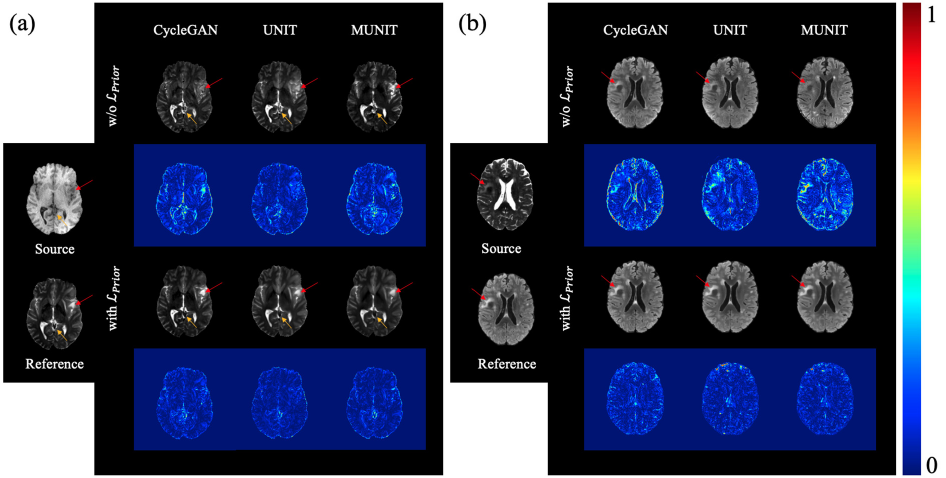}
    \caption{Qualitative comparison of CycleGAN and its variants incorporating the proposed prior loss $\mathcal{L}_{Prior}$ on the RA+MS datasets constructed by (a)BraTS2020 and (b)PDGM respectively.}
    \label{fig_12}  
  \end{figure*}
In addition, we also observed a surprising phenomenon.
Theoretically, when the training data is well-aligned, the performance of RegGAN and MITIA should be similar to Pix2Pix, as there is no misalignment error to interfere with the model optimization. However, in the quantitative results of Pix2Pix, RegGAN, and MITIA, we always have Pix2Pix $<$ RegGAN $<$ MITIA.
As shown in the regions indicated by the blue arrows in Figure \ref{fig_8} and Figure \ref{fig_9}, the result images of Pix2Pix and RegGAN always lack some fine structures, while MITIA can preserve these fine structures well.
A reasonable explanation for our results is that even in the well-aligned BraTS2020 and PDGM datasets, there are still misalignment errors that interfere with the model optimization.
Therefore, we examined the training images and the corresponding confidence matrix $W$. 
We found that in most cases, the MDet module could detect a small amount of unregistrable misalignment errors from the aligned training data (as shown in Figure \ref{fig_10}(a)).
In some special cases, such as when there are obvious artifacts in the images, the detected misalignment errors would increase markedly (as shown in Figure \ref{fig_10}(b)).

To further validate our argument, we also utilized a well-trained multi-modal error detector $D$ to detect and compare the misalignment errors before and after processing by MReg. 
The experiments were conducted on Paired and RA+MS training sets constructed by BraTS2020 and PDGM. 
The average misalignment error for each image pair is calculated and plotted as frequency distribution line graphs (Figure \ref{fig_11}).
It can be seen that the misalignment errors in the training data were reduced after processing by MReg, indicating the presence of registrable misalignment errors in both the Paired training set and the RA+MS training set. 
Moreover, we can see the blue lines in (b) and (d) have larger leftward shift compared to that in (a) and (c).
It is consistent with our data setting where RA+MS training set has notably more registrable misalignment errors than Paired training set. 
From the blue lines, it can be seen that misalignment errors in the Paired training data still exist after processed by MReg.
This result indicates that Paired training data still contains varying levels of unregistrable misalignment errors, which would interfere with model optimization.
This further demonstrates the difficulty of obtaining pixel-wise aligned data in medical scenarios and highlights the value of MITIA in practical applications.

\subsubsection{Performance of different models incorporating $\mathcal{L}_{Prior}$}
\begin{table}[!htb]
    \begin{center}
      \caption{Comparison of PSNR and SSIM for CycleGAN and its variants incorporating the proposed prior loss $\mathcal{L}_{Prior}$ on the RA+MS datasets constructed by BraTS2020 and PDGM respectively.}
      \label{table_5}
      \resizebox{\linewidth}{!}{
      \begin{tabular} {cccccc}
      \toprule[1.5pt]
                       & &     \multicolumn{2}{c}{BraTS2020}   &    \multicolumn{2}{c}{PDGM}    \\
      \cmidrule{2-6}
                       & $\mathcal{L}_{Prior}$ &     PSNR(dB)  &    SSIM(\%)   &    PSNR(dB)   &    SSIM(\%)  \\
      \midrule
                 \multirow{2}{*}{CycleGAN}  &            & 23.39$\pm$0.94 & 88.11$\pm$0.88 & 24.28$\pm$0.58 & 87.16$\pm$0.81 \\
                                            & \checkmark & 26.37$\pm$1.18 & 92.40$\pm$1.31 & 26.90$\pm$1.23 & 92.34$\pm$0.56 \\
                     \multirow{2}{*}{UNIT}  &            & 24.54$\pm$1.34 & 90.52$\pm$1.60 & 25.02$\pm$1.42 & 89.86$\pm$1.00 \\
                                            & \checkmark & 26.16$\pm$1.33 & 92.53$\pm$1.24 & 26.53$\pm$1.22 & 92.41$\pm$0.92 \\
                    \multirow{2}{*}{MUNIT}  &            & 22.20$\pm$1.01 & 86.77$\pm$1.09 & 22.96$\pm$1.01 & 86.29$\pm$1.22 \\
                                            & \checkmark & 24.89$\pm$1.21 & 91.26$\pm$1.11 & 25.63$\pm$1.18 & 89.68$\pm$1.03 \\
                  
      \bottomrule[1.5pt]
      \end{tabular}
      }
    \end{center}
  \end{table}

To validate the effectiveness and transferability of the proposed prior loss, we incorporated $\mathcal{L}_{Prior}$ as an additional term in the objective function of different unsupervised image-to-image translation models, including CycleGAN and its variants, UNIT and MUNIT.
In the new objective function obtained by introducing $\mathcal{L}_{Prior}$ into each model, the weight of $\mathcal{L}_{Prior}$, $\lambda_{Prior}$, was set to 30, while the weights of the other terms remained unchanged.
All experiments were conducted based on two RA+MS training sets constructed by BraTS2020 and PDGM.
Quantitative results are listed in Table \ref{table_5}, while representative images along with their residual maps compared to the reference images are presented in Figure \ref{fig_12}.
The residual maps in Figure \ref{fig_12} intuitively show that the error between the predicted images and the ground truth decreased noticeably after introducing $\mathcal{L}_{Prior}$.
This suggests that the guidance of pixel-level prior information enhanced the model's fidelity to image contents. From the regions indicated by the red and yellow arrows, it can be observed that the new methods incorporating $\mathcal{L}_{Prior}$ depict fine structures more accurately compared to the original unsupervised methods.
The quantitative results show that the performance metrics of CycleGAN and its variants improved to varying degrees after the introduction of $\mathcal{L}_{Prior}$.
CycleGAN-with-$\mathcal{L}_{Prior}$ achieved the highest PSNR scores of 26.37 dB and 26.90 dB. UNIT-with-$\mathcal{L}_{Prior}$ achieved the highest SSIM scores of 92.53\% and 92.41\%. Although MUNIT-with-$\mathcal{L}_{Prior}$ had lower performance metrics compared to the other methods, its PSNR and SSIM still improved by 2.69 dB and 4.49\% in BraTS2020 and by 2.67 dB and 3.39\% in PDGM, respectively, compared to MUNIT.
The above experimental results demonstrate that $\mathcal{L}_{Prior}$ can be integrated with various unsupervised models and effectively enhance their performance, highlighting the effectiveness and transferability of the proposed prior loss.

\subsection{Ablation study}
To validate the effectiveness of each module in MITIA, four experiments were conducted based on two RA+MS training sets constructed by BraTS2020 and PDGM, respectively. The experiment settings are as follows:

\begin{figure*}[!htb]
    \centering  
    \includegraphics[width=16cm]{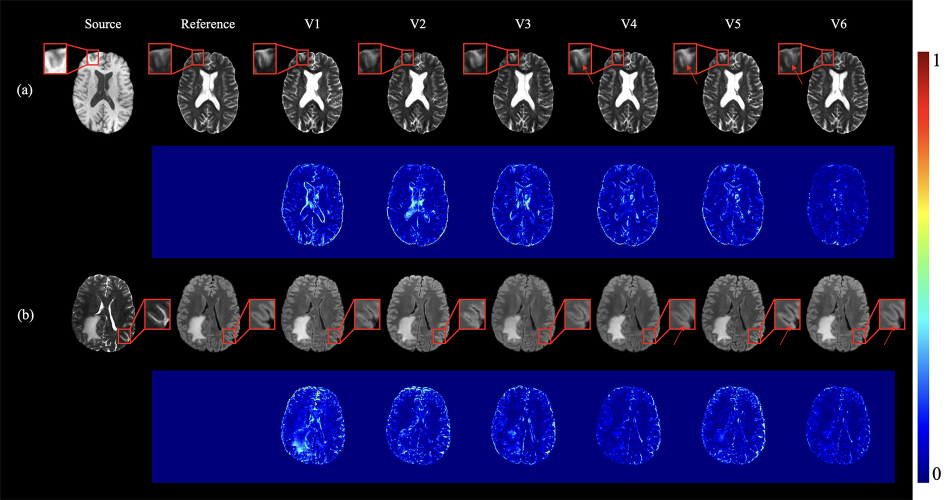}
    \caption{Qualitative comparison of the ablation study on the RA+MS datasets constructed by (a)BraTS2020 and (b)PDGM respectively.
    \label{fig_13}}  
  \end{figure*}

\begin{itemize}
  \item \textbf{V1}: CycleGAN model. This baseline model does not include the MReg and MDet modules, nor does it introduce any pixel-wise prior loss. It was trained only under the constraints of $\mathcal{L}_{Adv}$ and $\mathcal{L}_{Cyc}$.
  
  \item \textbf{V2}: Directly introducing pixel-wise prior loss to V1 without using the MReg and MDet modules. V2 is trained under the constraints of $\mathcal{L}_{Adv}$, $\mathcal{L}_{Cyc}$, and pixel-wise prior loss $\mathcal{L}_{V2}$ (Equation \ref{eq_15}). The weight of $\mathcal{L}_{V2}$ in the objective function is $\lambda_{V2}=30$.
  \begin{equation}
    \label{eq_15}
    \mathcal{L}_{V2} = \mathbb{E}_{x,\widetilde{y}}[|| G(x) - \widetilde{y}||_1]
  \end{equation}
  
  \item \textbf{V3}:  Introducing the fine registration model $R_F$ to V1, without using $R_C$ or the MDet module. V3 is trained under the constraints of $\mathcal{L}_{Adv}$, $\mathcal{L}_{Cyc}$, and pixel-wise prior loss $\mathcal{L}_{V3}$ (Equation \ref{eq_16}). The weight of $\mathcal{L}_{V3}$ in the objective function is $\lambda_{V3}=30$.
  \begin{equation}
    \label{eq_16}
    \mathcal{L}_{V2} = \mathbb{E}_{x,\widetilde{y}}[|| G(x \circ R_F(x, \tilde{y})) - \widetilde{y}||_1]
  \end{equation}
  
  \item \textbf{V4}: Introducing the MReg module to V1, and using $\mathcal{L}_{Adv}$, $\mathcal{L}_{Cyc}$, and the pixel-wise prior loss $\mathcal{L}_{V4}$ (Equation \ref{eq_17}) to constrain model optimization, where $x_f=x\circ\phi=x\circ(R_C(x,\widetilde{y})+R_F(x\circ R_C(x,\widetilde{y}),\widetilde{y}))$. The weight of $\mathcal{L}_{V4}$ in the objective function is $\lambda_{V4}=30$.
  \begin{equation}
    \label{eq_17}
    \mathcal{L}_{V4} = \mathbb{E}_{x,\widetilde{y}}[|| G(x_f) - \widetilde{y}||_1]
  \end{equation}
  
  \item \textbf{V5}: Introducing the MDet module to V1, and using $\mathcal{L}_{Adv}$, $\mathcal{L}_{Cyc}$, and the pixel-wise prior loss $\mathcal{L}_{V5}$ (Equation \ref{eq_18}) to constrain model optimization. The weight of $\mathcal{L}_{V5}$ in the objective function is $\lambda_{V5}=30$.
  \begin{equation}
    \label{eq_18}
    \mathcal{L}_{V5} = \mathbb{E}_{x,\widetilde{y}}[|| (G(x) - \widetilde{y}) \cdot Act( D(x,\widetilde{y}) )||_1]
  \end{equation}
  
  \item \textbf{V6}: The complete MITIA model, including the MReg, MDet, and Cycle modules. It uses the full objective as shown in Equation (\ref{eq_10}) during training.
\end{itemize}

\begin{table}[htb]
    \begin{center}
      \caption{Quantitative results of the ablation study on the RA+MS datasets constructed by BraTS2020 and PDGM respectively.
      \label{table_6}
      }
      \resizebox{\linewidth}{!}{
      \begin{tabular} {ccccccc}
      \toprule[1.5pt]
                       & & &     \multicolumn{2}{c}{BraTS2020}   &    \multicolumn{2}{c}{PDGM}     \\
      \cmidrule{2-7}
                       & MReg & MDet &     PSNR(dB)  &    SSIM(\%)   &    PSNR(dB)   &    SSIM(\%)  \\
      \midrule
                   V1  &&& 23.39$\pm$0.94 & 88.11$\pm$0.88 & 24.28$\pm$0.58 & 87.16$\pm$0.81 \\
                   V2  &&& 23.84$\pm$1.45 & 88.68$\pm$1.01 & 24.67$\pm$0.93 & 87.77$\pm$0.74 \\
                   V3  &only $R_F$&& 24.93$\pm$0.88 & 90.20$\pm$1.06 & 25.22$\pm$0.77 & 89.50$\pm$0.91 \\
                   V4  &\checkmark&& 25.52$\pm$0.95 & 91.04$\pm$1.25 & 25.89$\pm$1.09 & 91.31$\pm$1.10 \\
                   V5  &&\checkmark& 25.73$\pm$1.32 & 91.11$\pm$1.36 & 25.65$\pm$1.12 & 90.96$\pm$0.69 \\
          \textbf{V6}  &\checkmark&\checkmark& \textbf{26.37$\pm$1.18} & \textbf{92.40$\pm$1.31} & \textbf{26.90$\pm$1.23} & \textbf{92.34$\pm$0.56}  \\
      \bottomrule[1.5pt]
      \end{tabular}
      }
    \end{center}
\end{table}

Quantitative results are presented in Table \ref{table_6}, while representative images and their residual maps, compared to the reference images, are displayed in Figure \ref{fig_13}.
V2 shows a slight improvement in both metrics over V1, whereas V3, which incorporates $R_F$, demonstrates a notable enhancement in both metrics.
The improvement in V2 indicates that incorporating the prior loss term is helpful to improve model performance when there are misalignment errors in the training data.
The improvement in V3 suggests that combining the prior loss term with a registration network can provide more available pixel-level prior information for model optimization, resulting in a greater performance enhancement.
V4, which incorporates the coarse-to-fine registration module, MReg, shows a further improvement in performance metrics compared to V3.
This indicates that using coarse-to-fine registration is more effective in correcting registrable misalignment errors in the training data than using $R_F$ alone, thereby providing more reliable guidance for model optimization.
V5 also shows a notable improvement in performance metrics compared to V1, proving that the MDet module effectively prevents misalignment errors in the training data from interfering with model optimization.
V6 achieved the best quantitative results, with an improvement in PSNR by 2.98 dB and 2.62 dB, and SSIM by 4.29\% and 5.18\% compared to V1. This indicates that the MReg and MDet modules can synergistically improve model performance effectively.
From the residual maps in Figure \ref{fig_13}, it is visually evident that the predicted images in V1 and V2 exhibit visible errors, while the quality of predicted images in V4 and V5 shows noticeable improvement.
This suggests that both MReg and MDet can effectively enhance the quality of generated images.
Consistent with the quantitative results above, V6 exhibits the least amount of errors in its results, and it can depict the detailed structures in the images more accurately (as indicated by the red arrows in Figure \ref{fig_13}).

\section{Discussion and conclusion}
From the above experiment results, it can be seen that the proposed method achieves good performance on both well-aligned and misaligned datasets.
The success of our method is attributed to the following reasons. 
Firstly, our proposed method is based on the cycle-consistent GAN model, which has been proven to be powerful in generating images similar to the target images.
Secondly, the cascaded registration module, MReg, can effectively eliminate registrable misalignment errors, thus significantly increasing the correct pixel-level prior information in the training data.
Thirdly, the multi-modal misalignment error detection module, MDet, can exclude the remaining unregistrable misalignment errors in the training data, thereby providing more reliable guidance for model optimization.
Through extensive experiments, we have demonstrated that when facing different types and severity of misalignment errors, MITIA can generate images with more accurate content information and more details compared to other state-of-the-art methods, and it also has a significant advantage in PSNR and SSIM scores.
These results indicate that our proposed MITIA model has stronger anti-interference ability to misalignment errors in training data, benefiting from the introduction of the prior extraction network composed of the MReg and MDet modules.
In the ablation experiments, we demonstrated that both MReg and MDet are effective.
From the results of V2 and V3, we found that extracting correct prior information and removing incorrect prior information are equally important for improving the model's performance.
Since the idea of using registrable and unregistrable data in misaligned datasets for assisting unsupervised training is proposed for the first time, we have reason to believe that the performance of our constructed MReg and MDet modules in extracting registrable data and removing unregistrable data is not optimal.
Therefore, we can infer that by designing a more powerful multimodal medical image registration model and a more accurate multimodal misalignment error detection model to replace the MReg and MDet modules, further increasing the quantity and accuracy of extracted prior information, the performance of image-to-image translation models based on misaligned data can theoretically be further improved.
In addition, since pixel-wise prior constraints are applicable to most image-to-image translation models based on deep learning, the prior extraction network composed of the MReg and MDet modules in MITIA should have good transferability.
With the continuous emergence of powerful basic generative models in recent years (such as Diffusion\cite{b19,b32,b33,b34}, ViT\cite{b35,b36,b37,b38}, etc.), we believe that combining the prior extraction network with these basic generative models can achieve higher-quality image translation results.

In conclusion, we have proposed a novel GAN-based multi-modal medical image-to-image translation model termed MITIA, which achieves outstanding performance in multi-modal medical image-to-image translation tasks without relying on pixel-wise aligned training data. 
Through quantitative and qualitative analysis based on both well-aligned and misaligned datasets, we can conclude that MITIA achieves better performance and preserves more content information compared to other state-of-the-art methods.

\section*{Acknowledgment}
This work is supported in part by the National Key Research and Development Program of China (2022YFF0710800), and Jiangsu Provincial Key Research and Development Program (BE2021609).

\vspace{12pt}

\end{document}